\documentclass[sigconf]{acmart}
\AtBeginDocument{%
  }

\setcopyright{acmlicensed}
\copyrightyear{2026}
\acmYear{2026}
\acmDOI{XXXXXXX.XXXXXXX}
\acmConference[EASE Companion '26]{Proceedings of the 2026 Evaluation and Assessment in Software Engineering Companion}{June 09--12, 2026}{Glasgow, UK}
\acmBooktitle{Proceedings of the 2026 Evaluation and Assessment in Software Engineering Companion (EASE Companion '26), June 09--12, 2026, Glasgow, UK}

\acmISBN{978-1-4503-XXXX-X/2018/06}
\usepackage{xcolor}
\usepackage{soul}
\usepackage{array}
\usepackage{braket}

\usepackage{balance}

\begin{document}

\title{QubitQuest: Learning Quantum Computing through Mini-Games}

\author{Bella Hill}
\email{bella.hill@canterbury.ac.nz}
\orcid{0009-0009-4887-7323}
\affiliation{%
  \institution{University of Canterbury}
  \city{Christchurch}
  \country{New Zealand}
}

\author{Miguel Eh\'ecatl Morales-Trujillo}
\email{miguel.morales@canterbury.ac.nz}
\orcid{0000-0002-1706-1577}
\affiliation{%
  \institution{University of Canterbury}
  \city{Christchurch}
  \country{New Zealand}
}

\renewcommand{\shortauthors}{Hill \& Morales-Trujillo}

\begin{abstract}
Quantum Computing (QC) is often challenging for beginners due to its abstract concepts and mathematical foundations.
This paper explores the use of gamification to support the learning of introductory QC concepts.
To investigate this, \textit{QubitQuest} was developed as a set of three educational mini‑games designed to teach key QC topics: the Bloch sphere, entanglement, and quantum circuits.
The mini‑games aim to balance engagement, motivation, and learning while introducing concepts in small and focused units of progressive difficulty.
A two‑phase user study was conducted to evaluate the mini‑games.
In the first phase, a preliminary survey was conducted to gather information on learners' preferences and inform the design of the mini‑games.
In the second phase, participants played the mini‑games and completed pre‑ and post‑game questionnaires to assess their learning.
The results show that participants improved their understanding of introductory QC concepts after playing the mini‑games, with post‑game scores higher than pre‑game scores.
Those who completed more levels achieved higher post‑game scores, indicating that motivation and engagement influenced the learning outcomes.
These findings suggest that mini‑games may improve students' learning experience and outcomes when exposed to introductory QC concepts.
\end{abstract}




\maketitle

\section{Introduction}
\label{sec:intro} 

Quantum Computing (QC) is a computing paradigm that applies principles of quantum mechanics, such as superposition and entanglement, to computation \cite{Saini2025}. 
In QC, information is represented using quantum bits (qubits), which are manipulated using quantum gates. 
Unlike classical computing, where data is stored as bits with values of either `0' or `1', QC allows information to be processed in fundamentally different ways \cite{Weingartner23}.

QC's rapid development has shown its potential to solve problems more efficiently than classical computers \cite{Schneider24, Gamble19}.
The growth of this field is expected to increase the need for broader quantum literacy, particularly among STEM undergraduate students \cite{Xenakis23}. 
However, quantum mechanics concepts are complex and difficult to understand, even for experts in the field \cite{Evenbly24, Weingartner23}, creating opportunities to develop more effective ways to teach QC knowledge to students in an engaging and accessible manner \cite{Seegerer21}. 

Gamification, the use of game design elements in a non-gaming context \cite{Deterding11}, is a popular pedagogical strategy used by educators to increase student motivation and engagement \cite{Aibar24, Lee11, Nand19, Zeybek23}. 
Gamification has proven to be particularly effective in STEM subjects, with research showing that its use improves learning outcomes for students across a wide range of ages \cite{Baykal23, Dichev17, Linehan11, Zhang21,morales-trujillo2020,morales-trujillo2025}.

As a result, gamification has the potential to be a solution for teaching QC concepts, but it is necessary to determine which game elements and what game types are most suitable for the topic and target users \cite{Zeybek23}.
In addition, existing QC gamified resources often require a strong mathematical background or assume prior knowledge, making them less suitable for beginners.

This paper presents \textit{QubitQuest}, a set of three mini‑games designed to teach QC concepts to undergraduate students. 
The research examines how this gamified learning resource influences students’ performance, engagement, and motivation, guiding the following research questions:

\begin{itemize}
    \item \textbf{RQ1: How effective is gamification at improving students' learning outcomes in introductory QC concepts?}
    
    From a pedagogical perspective, this question aims to provide guidance on designing gamified learning experiences that improve students' motivation, engagement, and performance in introductory QC education.

    \item \textbf{RQ2: What is the relationship between students' motivation and engagement, and their performance?}
    
    From an empirical perspective, this question aims to examine how motivational and engagement factors are associated with measurable learning outcomes in a game-based learning environment.
\end{itemize}


The main contributions of this paper are: (i) the design and implementation of a gamified resource for teaching key QC topics through small learning units of progressive difficulty; (ii) the evaluation of the mini-games through a two-phase study with undergraduate students using pre- and post-game questionnaires to assess learning outcomes; and (iii) insights into how engagement and motivation influence participants’ completion of the game and how this relates to their learning outcomes.

The remainder of this paper is organised as follows. 
Section \ref{sec:preliminaries} introduces preliminary QC concepts, and Section \ref{sec:related-work} discusses related work. 
Section \ref{sec:solution} presents \textit{QubitQuest} and its mini-games, while Section \ref{sec:methodology} describes the study design. 
Section \ref{sec:results} presents the results, and Section \ref{sec:discussion} discusses them. 
Finally, Section \ref{sec:conclusion} concludes the paper and outlines future work.

\section{QC Preliminaries}
\label{sec:preliminaries} 

This section provides a brief overview of the fundamental quantum computing concepts used throughout this paper.

A \textit{qubit} is the basic unit of information in quantum computing \cite{Escanez23}. 
While a classical bit can only represent a `0' or a `1', a qubit can exist in a combination of both states at the same time with different probabilities \cite{Escanez23, Gamble19}. 
This property is known as \textit{superposition} and is one of the core concepts of QC.

\textit{Measurement} is the act of observing a qubit's state; when the quantum system is measured, the superposition of states collapses into either $\ket{0}$ or $\ket{1}$ \cite{Schneider24, Gamble19}. 
The \textit{Bloch sphere} is a visual representation of a single qubit's state, shown as a point on or inside a sphere. 
It helps illustrate how qubits can exist in many possible states between `0' and `1' by showing complex numbers in a three-dimensional plane~\cite{Escanez23}.

\textit{Entanglement} is another key concept of QC. 
Dubbed ``spooky action at a distance'' by Einstein, it is a special quantum connection between particles where the state of one instantly relates to the state of the other, even if they are far apart. 
When one particle is measured, information about the corresponding entangled particles is instantly known \cite{Gamble19, Schneider24}. 
\textit{Correlated entanglement} is a type of entanglement where the entangled particles always produce the same result when measured (e.g., if one is $\ket{0}$, the other is also $\ket{0}$). 
\textit{Anti-correlated entanglement} is another type of entanglement where measuring one particle always gives the opposite result of the other (e.g., if one is $\ket{0}$, the other is $\ket{1}$).

Any manipulation of a qubit to perform computations is called a \textit{quantum operation}. 
This includes entangling, measuring, or applying quantum gates to qubits. 
\textit{Quantum gates} are operations that change the state of qubits, similar to logic gates in classical computing. 
Each gate applies a specific transformation to one or more qubits, changing the probabilities in superposition \cite{Gamble19, Escanez23}. 
For example, the \textit{X gate}, or \textit{Pauli-X gate}, is the quantum equivalent of a classical NOT gate. 
It flips the state of a qubit (i.e., $\ket{0}$ becomes $\ket{1}$, and $\ket{1}$ becomes $\ket{0}$). 
The \textit{H gate}, or \textit{Hadamard gate}, is used to create or remove superposition. 
It transforms $\ket{0}$ into an equal probability of $\ket{0}$ and $\ket{1}$, and vice versa \cite{Liu23, Escanez23}.  

The gates can be mathematically represented by \textit{quantum matrices}. 
Each matrix defines how a gate transforms the state of a qubit or group of qubits, using linear algebra principles to describe quantum state changes \cite{Artner23}.

\textit{Decoherence} is the process by which a quantum system loses its quantum properties (like superposition or entanglement) due to interaction with its environment, leading to errors in computation \cite{Escanez23, Gamble19}. 
Reducing and minimising decoherence is one of the biggest challenges in QC. 
This is why quantum computers are kept at extremely low temperatures to reduce thermal noise and avoid decoherence, as they rely on preserving quantum properties to perform calculations and remain stable \cite{Schneider24}.

\section{Related Work}
\label{sec:related-work} 

This section provides an overview of existing games designed to teach QC concepts, identified through an ad hoc literature review using Scopus and Google Scholar. 
It also presents a detailed analysis of three selected games, whose strengths and limitations informed the design of \textit{QubitQuest}.

\subsection{Overview of Existing QC Games}

Most games designed to teach QC concepts target students from intermediate school to the undergraduate level. 
Several games are aimed at younger or early-stage learners, such as \textit{Quander} \cite{Liu23} and \textit{Entanglion} \cite{Weisz18}, which were designed for intermediate school students and above. 
Other games, like \textit{Qubit: The Game} \cite{Escanez23} and \textit{Quantum Odyssey} \cite{Nita21}, target high school students and above, while games such as \textit{Particle in a Box} \cite{Anupam18} and \textit{QRogue} \cite{Artner23} are mainly intended for undergraduate students.

Some games were not designed with a specific age group in mind, but instead aim to introduce QC concepts to learners with no prior background in the field. 
Examples include \textit{The Qubit Factory} \cite{Evenbly24} and the \textit{Virtual Lab by Quantum Flytrap} \cite{Jankiewicz22, Migdal22}, which focus on exploratory learning and conceptual understanding.

In terms of format, many QC educational games are web-based puzzle games \cite{Seskir22}, such as \textit{Quander} and \textit{The Qubit Factory}. 
Other formats include physical board games (e.g., \textit{Entanglion}, \textit{Qubit: The Game}), online platform-style games (e.g., \textit{Particle in a Box}), and simulator-based environments (e.g., \textit{Virtual Lab by Quantum Flytrap}). 
This variety shows that QC concepts can be presented through distinct interaction styles, each with its own pros and cons.

Not all identified games could be examined in detail. 
\textit{Qubit: The Game} was unavailable to play at the time of this study, \textit{Particle in a Box} provided only a short demo version, and \textit{Entanglion} is a multiplayer board game, making direct comparison difficult. 
Games were excluded when they were unavailable, not playable in full, or not supported by published research.

The main characteristics, strengths, and limitations of these games are summarised in Table~\ref{tbl:games-summary}.
Based on these observations, three games were selected as the most relevant for this research and are described in more detail in the following subsections.

\subsection{Quander}
\textit{Quander} is a series of mini-games designed to teach introductory QC concepts in a \textit{``fun, uncomplicated, and engaging manner''} \cite{Liu23}. 
The game has a supernatural theme, with a vampiric main character tasked by her ``genius scientist'' cat to collect parts for a quantum computer. 
A narrative aspect, with different characters in each mini-game, and an overall goal are the game's core elements. 

\textit{Qupcakery}, one of \textit{Quander}'s mini-games, is a cooking-style puzzle game where the player uses quantum gates to change \textit{qupcakes} (quantum cupcakes) to be either vanilla-flavoured ($\ket{0}$) or chocolate-flavoured ($\ket{1}$). 
The main QC concepts taught in \textit{Qupcakery} are the quantum states, measurement, qubits, quantum operations, and entanglement.

Another mini-game in \textit{Quander} is \textit{TwinTanglement}, where two zombies move simultaneously through a maze.
The player must switch between twins in control to get them both out of the maze. 
This game teaches the concepts of entanglement, quantum communication, qubits, and correlation.

The remaining three mini-games are \textit{QueueBits}, a Connect 4-inspired game covering probability, superposition, and measurement; \textit{Buried Treasure}, a treasure hunting game covering measurement and quantum sensing; and \textit{Tangle's Lair}, a quantum circuit simplification game covering quantum gates and quantum circuits.

While \textit{Quander}'s target demographic is intermediate to early high school-aged children, it focuses on minority groups who are underrepresented in higher STEM education and therefore have fewer opportunities to learn QC concepts \cite{Liu23}. 

The main component of \textit{Quander} that will be utilised is the use of different mini-games, thematically linked, each teaching a different QC concept.

\subsection{QRogue}
\textit{QRogue} is a text-based command-line interface puzzle game designed to teach QC concepts in a ``playful'' yet mathematical way, directly showing the matrix-vector multiplications that are involved in quantum operations \cite{Artner23}. 
The game has a robot character traversing different rooms to battle enemies. 
The robot is called a Qubot, and the enemies are QC puzzles that the player must solve. 

These puzzles start as simple quantum circuit puzzles and become progressively harder, introducing more concepts as the game advances. 
Throughout the rooms, the player will be shown information pop-ups that explain new QC concepts they need in order to solve the puzzles. 

The main QC concepts \textit{QRogue} aims to teach are superposition, measurement, entanglement, and quantum gate operations, with a specific focus on the underlying mathematical operations \cite{Artner23}. 
The latter of these relates closely to the topic of “qubits as matrices”, which is worth noting, as it is a difficult concept to teach, and most QC games do not attempt to do it explicitly. 

\subsection{Quantum Odyssey}
\textit{Quantum Odyssey} is an immersive story-based puzzle game that shows visual representations of quantum gates rather than teaching them mathematically. 
The game visually illustrates qubit states using colourful balls that fall along lines representing quantum circuits. 
With a single qubit, there are two lines, each representing a state. 
The first line corresponds to the $\ket{0}$ state, and the second to the $\ket{1}$ state. 

The ball starts in a given input state, going down the line to the output state, and the user applies quantum gates to change the destination of the ball, which visually shows the output state changing by connecting the different lines or changing the ball's colour. 
For example, if the user places an X gate on a single qubit whose state starts at $\ket{0}$, then the two lines of the circuit are crossed, allowing the ball to physically transition from $\ket{0}$ to $\ket{1}$. If there are two qubits, then four lines are possible ($\ket{00}$, $\ket{01}$, $\ket{10}$, $\ket{11}$). 

Colours are used to represent negative and imaginary numbers, and superposition is shown by the ball splitting into smaller balls and therefore being able to end up in multiple destinations. 
The game has progressive levels with different puzzles, initially focusing on individual concepts and then combining them as the user advances further. 

\begin{table}[!ht]
\scriptsize
\caption{Summary of Educational QC Games}
\label{tbl:games-summary}
\begin{tabular}{p{1cm} p{1.2cm} p{1.1cm} p{3.7cm}}
\hline
\textbf{Game} & \textbf{Target Audience} & \textbf{Format} & \textbf{Main Characteristics} \\
\hline
\hline

Quander \cite{Liu23}
& Intermediate school and above
& Web-based mini-games
& Decomposes QC topics into small learning units; good visual design; concepts are simplified; advanced topics are not covered \\ \hline

Entanglion \cite{Weisz18}
& Intermediate school and above
& Physical board game
& Focuses on entanglement concepts; multiplayer format makes direct comparison with digital games difficult \\ \hline

Qubit: The Game \cite{Escanez23}
& High school and above
& Physical board game
& Designed to introduce QC concepts through cards representing events and qubits; unavailable to play at the time of this study \\ \hline

Quantum Odyssey \cite{Nita21}
& High school and above
& Digital game
& Uses visual metaphors to explain QC concepts; supported by existing literature; key topics are missing; the teaching is too abstract \\ \hline

Particle in a Box \cite{Anupam18}
& Undergraduate
& Online platform game
& Focuses on quantum mechanics concepts; only a short demo version was available \\ \hline

QRogue \cite{Artner23}
& Undergraduate
& Web-based game
& Covers qubits as matrices and uses increasing difficulty levels; limited scope; outdated look and feel; difficult to use \\ \hline

The Qubit Factory \cite{Evenbly24}
& General
& Web-based puzzle game
& Aims to introduce QC concepts without prerequisites; unclear instructions causing low engagement \\ \hline

Virtual Lab by Quantum Flytrap \cite{Jankiewicz22}
& General
& Simulator-based environment
& Interactive QC simulations for exploration; less game-like and not strongly focused on engagement \\ 
\hline
\hline

\end{tabular}
\end{table}

\section{QubitQuest: the Solution}
\label{sec:solution} 

\textit{QubitQuest} is a collection of mini-games aiming to keep the learning experience simple and the progress gradual. 
The mini-games are linked by an overarching cat theme, inspired by the well-known quantum mechanics experiment: the Schr{\"o}dinger's Cat. 

\begin{figure}[!ht]
 \centering    
 \includegraphics[width=1\linewidth]{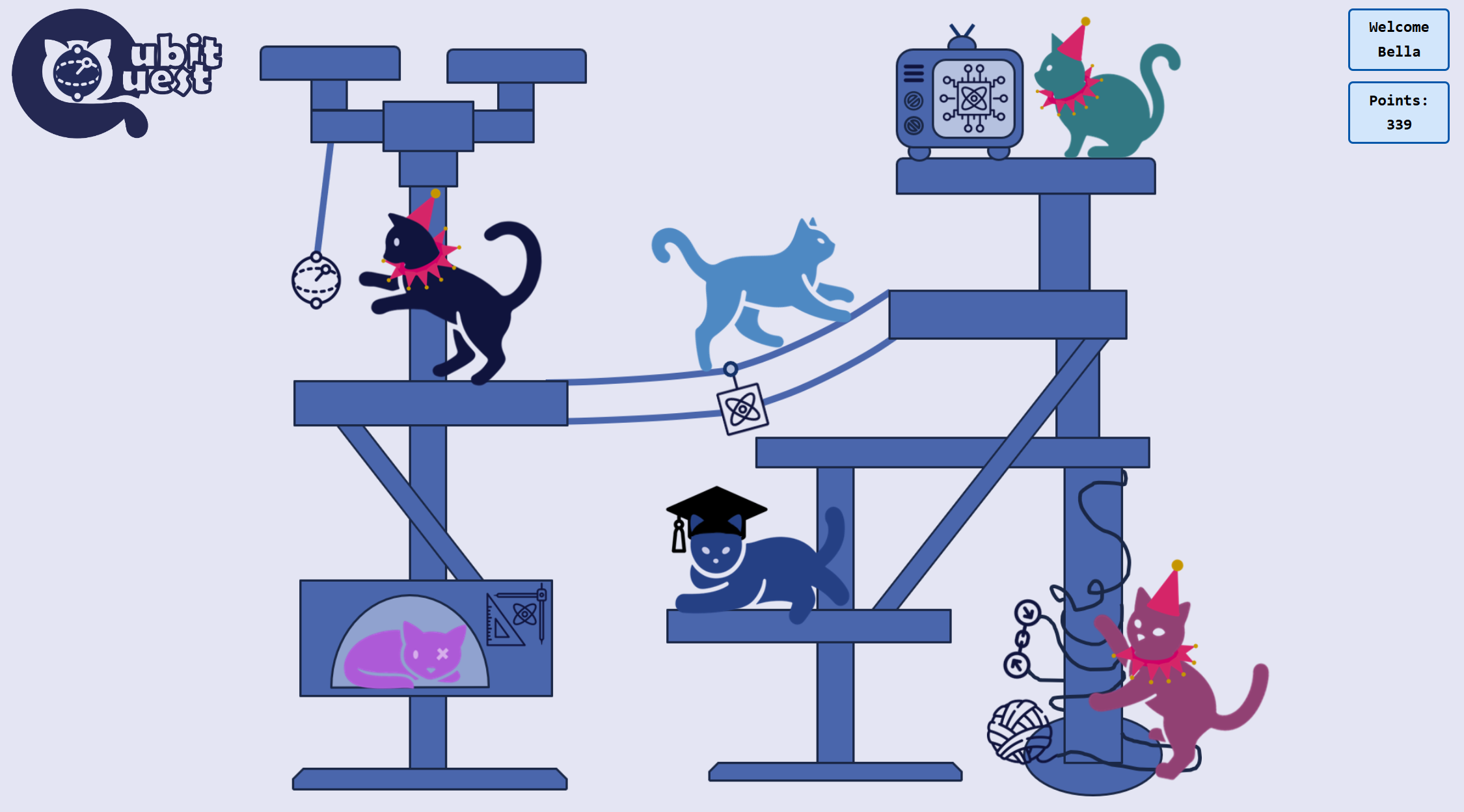}
   \caption{QubitQuest Main Page}
    \label{fig:main-page}
\end{figure}

The home screen displays a cat tree with several cats, each representing a different mini-game, and a ``special'' cat that links to the in-game quizzes to test the user's knowledge on QC (see Figure \ref{fig:main-page}). 
Particular focus was placed on aesthetic appeal and engagement, as well as on preferences indicated by participants in the preliminary survey (see Section \ref{sec:methodology}).

The three mini-games chosen in \textit{QubitQuest} are the Bloch sphere, entanglement, and quantum circuits.
These topics are core introductory concepts in QC and are commonly introduced early in undergraduate QC courses. 
Together, they cover conceptual and operational aspects of QC. 
The Bloch sphere was chosen because it provides an intuitive visual representation of a single qubit and its states, making it well-suited for introducing superposition and state manipulation. 
Entanglement was included as it is a fundamental concept that introduces the non-classical behaviour of multi-qubit systems and is often difficult for students to understand without visual support \cite{Nita21, Nita23, Jankiewicz22}. 
Quantum circuits were chosen to expose students to a more formal and practical representation of quantum computation, linking abstract concepts to how quantum algorithms are constructed. 
By separating these topics into individual mini-games, each concept could be introduced and explored independently, reducing cognitive load and allowing students to focus on one concept at a time.
The mini-game architecture allows the addition of new mini-games covering new QC topics, making the game's architecture scalable and open to extensions.

\subsection{The Mini-Games}

\begin{figure}[!ht]
    \centering
    \includegraphics[width=1\linewidth]{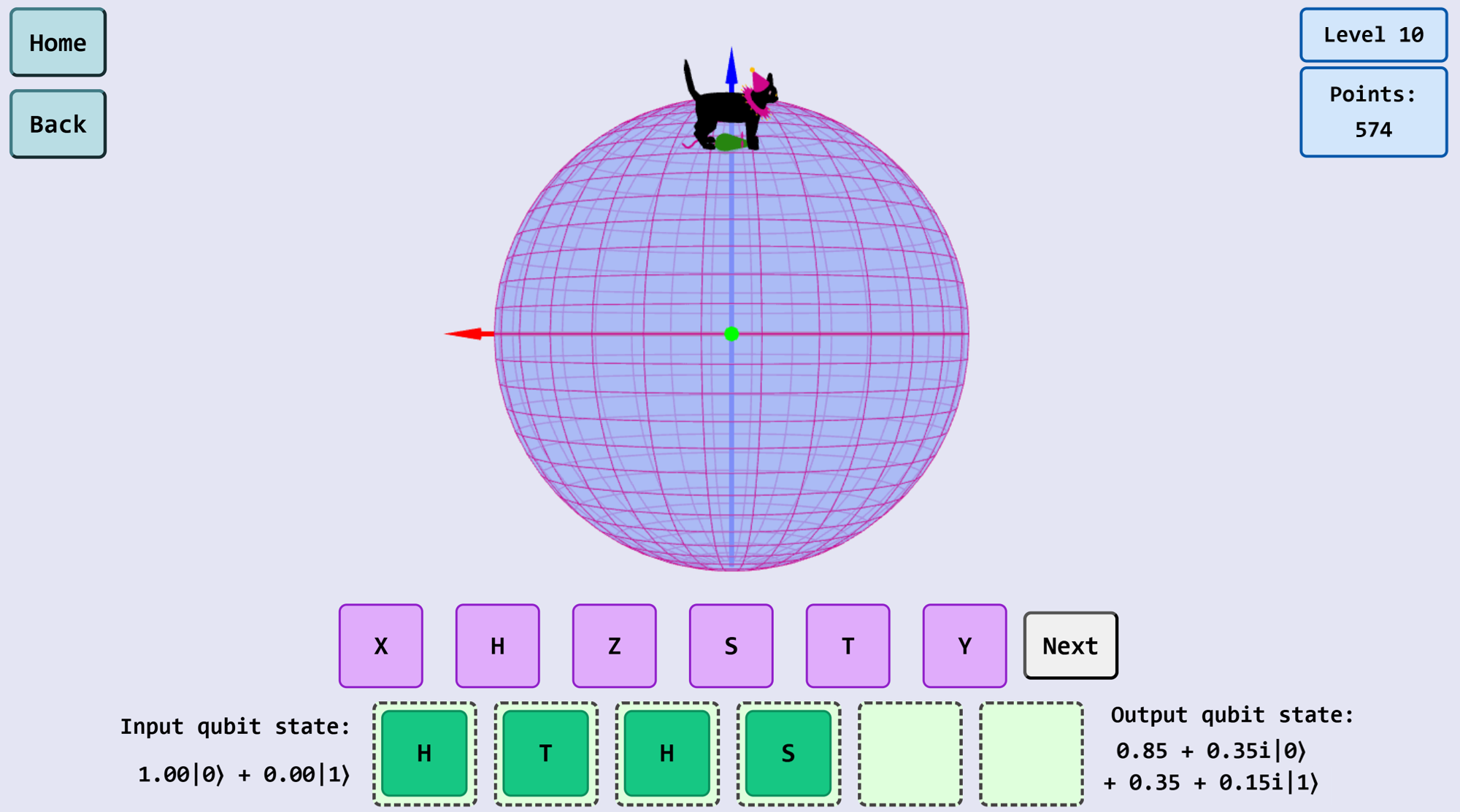}
    \caption{Bloch Sphere Mini-Game}
    \label{fig:bloch-sphere-page}
\end{figure}

\textbf{Bloch Sphere Mini-Game:} 
The focus of the Bloch sphere mini-game is to introduce quantum gates for a single qubit, visually represented on the Bloch sphere (see Figure \ref{fig:bloch-sphere-page}). 
The main QC concepts in this mini-game are superposition, qubits, quantum gates, and the Bloch sphere.
To gamify this, a 3D cat model represents the current qubit state on the Bloch sphere, and a 3D mouse toy represents the target qubit state. 

The aim of each level is to apply a combination of quantum gates to get the cat to reach the mouse on the Bloch sphere. 
Pop-up explanations and hints clarify quantum gates and the concept of superposition throughout the game. 
Each quantum gate also has a tooltip that provides a brief description and explains its relationship to the movement on the Bloch sphere. 

\begin{figure}[!ht]
    \centering
    \includegraphics[width=1\linewidth]{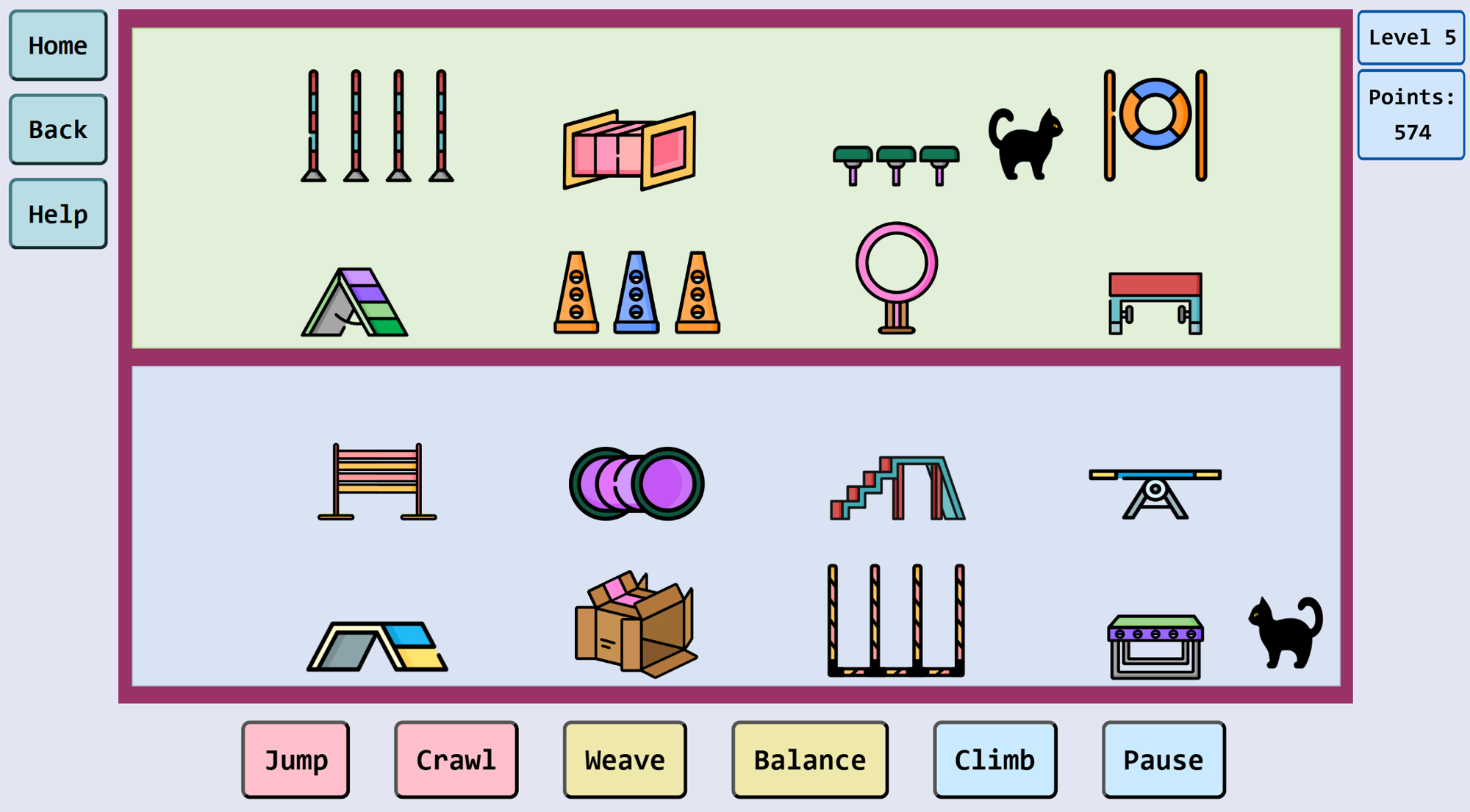}
    \caption{Entanglement Mini-Game}
    \label{fig:entanglement-page}
\end{figure}

\textbf{Entanglement Mini-Game:} 
The main QC concepts in the entanglement mini-game are entanglement, measurement, and decoherence.
This mini game involves a cat agility course with a series of obstacles for cats to navigate (see Figure \ref{fig:entanglement-page}).
There are two agility courses with two cats (one cat per agility course), and the cats are entangled. 
This means the user controls one cat, and the other cat performs the same actions as the first. 
The user is not able to switch control between cats. 
The user is required to navigate both cats through their respective obstacle courses, limiting the number of wrong moves, and gaining points when both cats successfully get through an obstacle at the same time. 

Several levels into the mini-game, the concept of decoherence is introduced; making wrong moves increases decoherence, while making synced moves decreases it. 
After six levels, the cats' behaviour changes to anti-correlated entanglement, where they perform exactly the opposite actions defined as follows: Jump--Crawl, Balance--Weave, and Climb--Pause.


The actual gameplay does not directly explain QC concepts, but again, explanatory pop-ups for each relevant level do. 

\begin{figure}[!ht]
    \centering
    \includegraphics[width=1\linewidth]{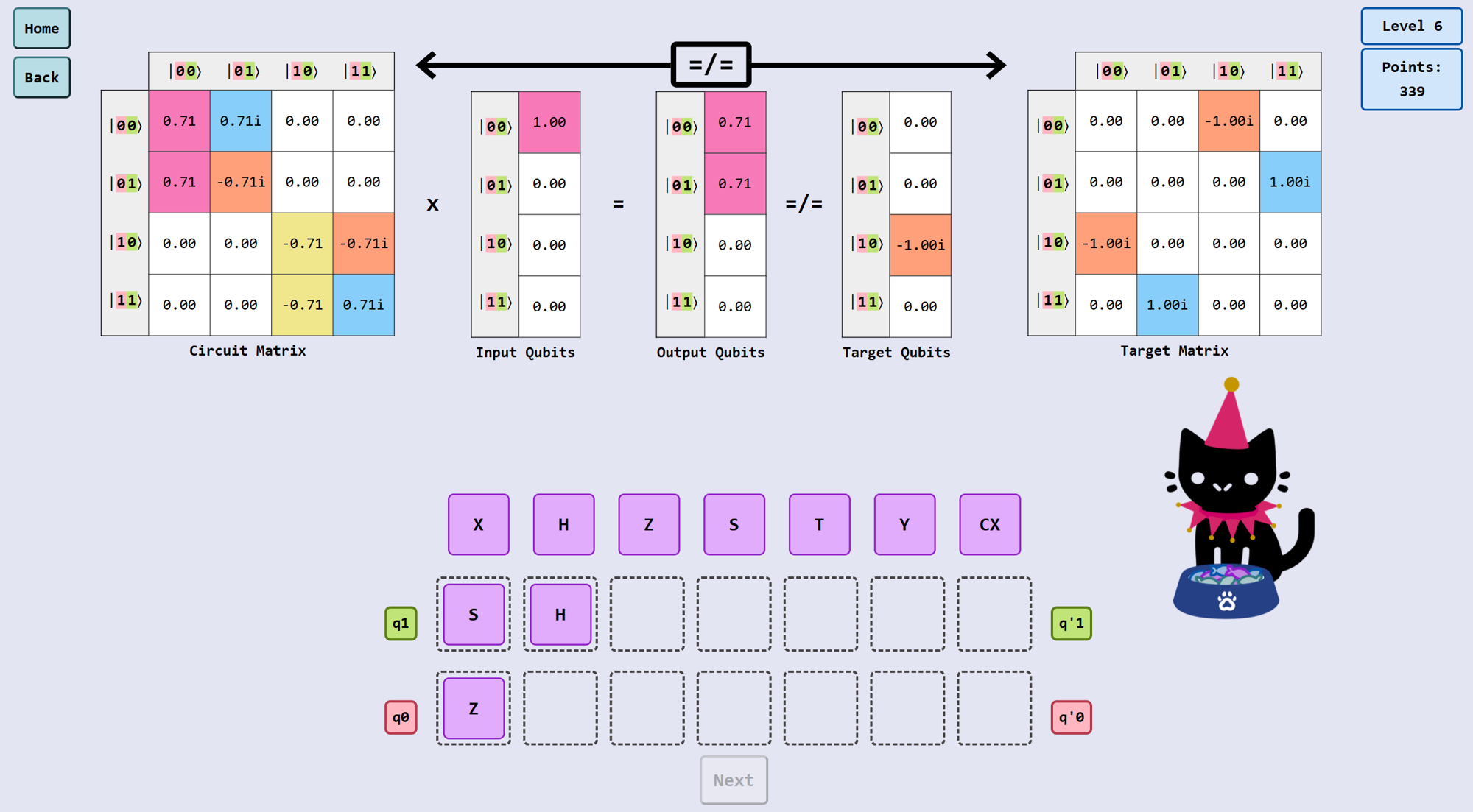}
    \caption{Quantum Circuits Mini-Game}
    \label{fig:quantum-circuits-page}
\end{figure}

\textbf{Quantum Circuits Mini-Game:} 
The quantum circuits mini-game is designed to teach the application of quantum gates to two qubits.
This is visualised using a matrix representation of a two-qubit circuit (see Figure \ref{fig:quantum-circuits-page}). 
This mini-game was inspired by \textit{QRogue}, which takes a similar approach using a matrix representation, but has been improved and further gamified. 

The main QC concepts in this mini-game are qubits, quantum gates, quantum circuits, and qubits as matrices.
To help with the visualisation, colour coding for the matrices was implemented to make it clearer what is happening at any moment. 
Positive real numbers are coloured pink, negative real numbers are yellow, positive imaginary numbers are blue, and negative imaginary numbers are orange. 
Combinations of two of these are shown as a gradient. 

Additionally, the target circuit matrix and the target qubit matrix are displayed, as in this version of the game, the user must match both. 
This is necessary because multiple different circuit matrices can generate the same output qubit matrix. 

To more explicitly gamify it and link it to the cat theme, a cat with a bowl of nine fish was added. 
Each time the user removes a gate from the circuit, the cat loses both a fish and a point from the 10 points available. 
After each three fish lost, it loses part of its outfit, finally becoming \textit{`sad'} and \textit{`starving'}, prompting the user to retry the level. 
This penalty is applied only from level two onwards, allowing users to play around with the gates. 

This mini-game uses the same quantum gates as the Bloch sphere game, but adds the CNOT gate, which requires two qubits. 
Similar to the other mini-games, gates also have tooltips that, in addition to their descriptions, also show the gate's effect on the matrix. 

\subsection{Game Design Decisions}

The main game elements in the final version were levels with increasing difficulty, a point-based system, and a rewards feature that unlocks a little jester outfit for the relevant cat on the main page.

Although narrative/storyline was in the initial design of the game, it was low-ranked in terms of user engagement preferences. 
Instead, the mini-games were linked together by the shared cat character and theme. 
User preferences also did not prioritise implementing badges,  achievements, or leaderboards.

Each mini-game has 12 levels, with the first level being an easy introductory level. 
The other levels become progressively harder yet remain achievable given what has been taught in the game. 
Each mini-game level also awards the user points, which accumulate across all mini-games and are displayed on the main page in the upper-right corner, along with the user's nickname (see Figure \ref{fig:main-page}).
Replaying an already completed level awards the user half the points they would have scored on first play, encouraging them to complete additional levels to earn more points.

The in-game quizzes do not award points to encourage users to complete them purely out of a desire to see how well they have learned, but they do have a high score out of 10 and show how many attempts they have made for each quiz.
This provides visual feedback to the user that they have completed the quiz and encourages them to replay it to achieve a perfect score.

The quantum circuits mini-game is considered the most complex in terms of QC knowledge required and, as such, is locked to the user until they have completed six levels of the Bloch sphere mini-game. 
This serves two purposes: to prevent users from starting with the most intimidating mini-game and being likely to get frustrated, and to build on the knowledge of the Bloch sphere mini-game, using the same quantum gates but now with two qubits. 
This was also influenced by the preliminary survey results, which indicated a desire for a progressive order of mini-games, rather than completion in any order. 

\textit{QubitQuest} is available to play online \cite{Hill25}, and further details on the game mechanics and rules are provided there.

\section{Methodology}
\label{sec:methodology} 

This section describes the design of the two-phase empirical study, the instruments, and the procedures for data collection and analysis.

\subsection{Study Design and Participants}
This study was composed of two phases. Phase One involved gathering students' preferences regarding their learning and game elements. 
For this purpose, participants answered a survey which results were used to make decisions on the mini-games' design.

Phase Two evaluated the effectiveness of the mini-games for teaching introductory QC concepts. 
Participants completed a pre-game questionnaire, interacted with the game, and then completed a post-game questionnaire. 
The same knowledge-based questions were used in both questionnaires to allow direct comparison of learning outcomes.

Participants were undergraduate Computer Science (CS) or Software Engineering (SE) students who could potentially take a QC course. Recruitment was conducted through the preliminary survey distributed to students enrolled in CS/SE courses. 

Participation in the study was voluntary, and no incentives were provided. 
Responses were anonymised, and participants were able to withdraw at any time.
This study was approved by the Human Research Ethics Committee (HREC) of the University of Canterbury.

\subsection{Phase One: Designing the Game}
Phase One involved a preliminary online survey conducted prior to the development of \textit{QubitQuest}. 
The survey, based on \cite{Dichev17,Linehan11}, included questions about learning preferences, game elements, aesthetics, and features that could help make the mini-games engaging, motivating, and effective when learning QC concepts (see Table \ref{tbl:preliminary-survey-questions}).
It was distributed to undergraduate CS/SE students at the University of Canterbury (New Zealand).
At the end of the survey, participants were given the option to provide their email address to be contacted for Phase Two of the study. 
This served as the main recruitment method for Phase Two.

The survey results (see Section \ref{sec:phaseone-results}) informed the main game design decisions.
Participants ranked visualisations and immediate feedback as the most helpful features for understanding complex concepts; consequently, all mini-games include these.
Levels with increasing difficulty and short play sessions preferences led to the 12-level structure per mini-game intended to be completed within a few minutes.
High ratings for points, rewards, and aesthetics resulted in a scoring system, cosmetic rewards, and a consistent visual theme. 
Low interest in narrative and competition led to their exclusion.

Regarding game mechanics, each mini-game follows a sequential level progression, i.e., completing a level unlocks the next. 
Failure results in retry without restriction. 
Penalty systems (e.g., point loss in circuits) favour efficient solutions but do not block progress. 
The only gating mechanism is unlocking the quantum circuits mini-game after completing six Bloch sphere levels. 
Otherwise, players may freely replay levels and explore mini-games.

\textit{QubitQuest} was developed using Java, JavaScript, and TypeScript.
The 3D graphics for the Bloch sphere game were implemented using Three.js with TypeScript and Webpack.

\begin{table}[!ht]
\scriptsize
\caption{Preliminary Survey Questions}
\label{tbl:preliminary-survey-questions}
\begin{tabular}{p{6.25cm} >{\raggedright\arraybackslash}p{1.25cm}}
\hline
\textbf{Question} & \textbf{Answer} \\
\hline
\hline
How comfortable are you with quantum computing concepts?
& Likert scale \\ \hline

Do you prefer games that require concepts to be learned progressively in a specific order, or do you like the flexibility to choose the sequence of mini-games (or no preference)?
& Single-choice \\ \hline

Would you find a quiz or challenge helpful before unlocking a new mini-game to reinforce concepts?
& Likert scale \\ \hline

How important are the following game elements to keep you engaged? a) Narrative/Storyline; b) Leaderboards; c) Levels with increasing difficulty; d) Points/scoring system; e) Badges/achievements; f) Progress Bars; g) Quests or challenges; h) Rewards (e.g., unlocking new features or content); and i) Aesthetics/visual appeal
& Likert scale for each element \\ \hline

What makes it easier for you to understand complex concepts in a game setting? a) Clear visualisations; b) Step-by-step instructions; c) Immediate feedback on actions; and d) Challenges that test understanding
& Ranked order \\ \hline

Do you prefer the game to: a) Focus primarily on teaching concepts in a structured way; b) Balance teaching with engaging gameplay; or c) Prioritise gameplay over teaching?
&  Single-choice \\ \hline

How long would you ideally like to spend on a single mini-game level?
& Single-choice \\ \hline

What motivates you to complete levels in educational games? a) Unlocking new content; b) Achieving high scores; c) Gaining a better understanding of the subject; and d) Competing with others
& Multiple-choice \\ \hline

What challenges do you typically face when learning new topics through games, and how could this game address those challenges?
& Open-ended \\ \hline

Are there any specific game mechanics or features you would like to see in a quantum computing game?
& Open-ended \\ \hline

Any additional feedback or suggestions for the design of these quantum computing mini-games?
& Open-ended \\ \hline
\hline

\end{tabular}
\end{table}

\subsection{Phase Two: Playing the Game}

This phase consisted of three sequential activities: (i) a pre-game questionnaire, (ii) gameplay with optional in-game quizzes, and (iii) a post-game questionnaire. 
Participants were asked to complete the questionnaires, while gameplay duration and quiz participation were partially self-directed.

\textbf{Pre-Game Questionnaire.} 
Participants first completed a pre-game questionnaire to assess their knowledge about QC. 
This questionnaire consisted of 10 multiple-choice questions on introductory QC concepts (see Table \ref{tbl:quiz-questions}). 
The questions were based on the knowledge taught in the game and the material developed in \cite{Usaola25}.

Each question included five answer options: one correct answer, three plausible incorrect answers, and an ``I don't know'' option. 
The ``I don't know'' option was included to discourage guessing and was pre-selected by default. 
Correct answers were not shown to participants. 
Access to the game was provided only after completion of the pre-game questionnaire to ensure all participants completed it before gameplay.

\begin{table}[!ht]
\scriptsize
\caption{Pre- and Post-Game QC Quiz Questions}
\label{tbl:quiz-questions}
\begin{tabular}{p{1.9cm} p{5.6cm}}
\hline
\textbf{Question} & \textbf{Answers (correct answer in bold)}  \\ \hline
\hline
Which of the following best describes superposition in quantum computing? 
&
a) A qubit being either $\ket{0}$ or $\ket{1}$ with certainty; 
\textbf{b) A qubit existing in a combination of $\ket{0}$ and $\ket{1}$ states simultaneously}; 
c) A qubit switching rapidly between $\ket{0}$ and $\ket{1}$; or 
d) A qubit being entangled with another qubit
\\ \hline

Which of the following best describes quantum entanglement? 
&
a) A single qubit being in multiple states at once; 
b) A method for reducing noise in quantum systems; 
c) A process of measuring qubits without collapse; or
\textbf{d) Two qubits whose states are correlated regardless of distance}
\\ \hline

Which of the following best describes a quantum gate? 
&
a) A physical barrier that traps qubits; 
b) A cooling device for superconductors; 
\textbf{c) A mathematical operation that changes a qubit's state}; or d) A measurement tool for qubits
\\ \hline

What is the function of the Hadamard (H) gate? 
&
a) It entangles two qubits together; 
\textbf{b) It creates a superposition of $\ket{0}$ and $\ket{1}$ from a single qubit}; 
c) It measures a qubit without collapse; or 
d) It prevents decoherence during computation
\\ \hline

Which of the following statements about the Z (Pauli-Z) gate is correct? 
&
\textbf{a) It changes the relative phase of the qubit without flipping its state}; 
b) It flips the qubit between $\ket{0}$ and $\ket{1}$; 
c) It places the qubit on the equator; or
d) It leaves the qubit unchanged
\\ \hline

What is a quantum circuit? 
&
a) A physical wire system that stores qubits;
b) A cooling system for superconducting processors; 
c) A measurement tool for qubits; or 
\textbf{d) A sequence of quantum gates applied to qubits}
\\ \hline

Why do quantum computers need to be kept at extremely cold temperatures? 
&
a) To prevent qubits from overheating and breaking; 
b) To reduce interference from surrounding electronic devices; 
\textbf{c) To minimise thermal noise and decoherence}; or 
d) To stop qubits from losing their spin orientation
\\ \hline

What is quantum decoherence? 
&
a) The process of a qubit splitting into two smaller qubits;
\textbf{b) The loss of quantum behaviour due to environmental interaction}; 
c) The ability of qubits to exist in multiple states; or
d) The collapse of a quantum system into entanglement
\\ \hline

What does it mean when a qubit is measured? 
&
\textbf{a) The qubit collapses into either $\ket{0}$ or $\ket{1}$}; 
b) The qubit becomes entangled with another qubit; 
c) The qubit stays in superposition; or 
d) The qubit cannot be used again
\\ \hline

What is the purpose of a controlled gate, like CNOT? 
&
a) To measure qubits without collapsing them; 
b) To keep qubits in superposition indefinitely; 
c) To erase errors in a quantum circuit; or 
\textbf{d) To apply an operation on one qubit depending on the state of another}
\\ \hline
\hline
\end{tabular}
\end{table}

\begin{table}[!ht]
\scriptsize
\caption{Post-Game Questionnaire}
\label{tbl:post-game-questionnaire}
\begin{tabular}{p{6.1cm} >{\raggedright\arraybackslash}p{1.4cm}}
\hline
\textbf{Question} & \textbf{Answer} \\
\hline
\hline
How much do you think the game contributed to your learning of introductory quantum computing concepts?
&  Likert scale \\ \hline

How long in total did you spend playing the game (in minutes)?
& Open-ended \\ \hline

Which of the mini-games did you play?
& Multiple-choice \\ \hline

How many levels per game did you play?
& [0-12] sliding scale per game \\ \hline

Please rank the games in order of personal preference (not how well you think it taught you, unless that is the biggest factor for your personal preference).
& Ranked order \\ \hline

Please rank the games in order of how effective at teaching you believe it was.
& Ranked order \\ \hline

Please rank the games in order of how engaging you found them to be.
& Ranked order \\ \hline

Please rank the games in order of how much you felt they motivated you to learn.
& Ranked order \\ \hline

Please select the option which is most relevant to each row: (see Figure \ref{fig:engagement-questions})
& Likert scale for each option \\ \hline

From the following feelings, tick those that you experienced while playing the games. Anxious, Amused, Sad, Angry, Happy, Lost, Interested, Attentive, Useful, Positive, Satisfied, Calmed, Bored, Immersed, Challenged, Confident, Competent, Stressed
& Multiple-choice \\ \hline

Did you complete any of the in-game quizzes? If so, please write which quizzes.
& Single-choice/open-ended \\ \hline

Did you find the in-game quizzes useful? (Please check N/A if you didn't complete any)
& Single-choice \\ \hline

Did you find the questions in the in-game quizzes to be an appropriate difficulty level for what the games taught you? (Please check N/A if you didn't complete any)
& Single-choice \\ \hline

What was your favourite thing(s) about the game?
& Open-ended \\ \hline

What was your least favourite thing(s) about the game?
& Open-ended \\ \hline

Do you have any feedback or suggestions for the game for improvements or new features?
& Open-ended \\ \hline

\hline
\end{tabular}
\end{table}

\textbf{Gameplay and In-game Quizzes.} 
After completing the pre-game questionnaire, participants were directed to play the mini-games and were encouraged to spend 10-20 minutes per mini-game and to attempt the in-game quizzes, although these were optional.

Three in-game quizzes were included, one for each mini-game and consisting of 10 multiple-choice questions. 
These quizzes could be replayed to achieve a high score and provided immediate feedback indicating whether answers were correct or incorrect. 
The in-game quiz questions were related to the content of the mini-games, with some general QC concept questions. 
No overlap between the in-game quiz questions and the pre-game or post-game questionnaire questions existed.

\textbf{Post-Game Questionnaire.} 
After playing the game, participants completed a post-game questionnaire (see Table \ref{tbl:post-game-questionnaire}). 
This questionnaire included the same 10 multiple-choice QC questions as the pre-game questionnaire to allow comparison of learning outcomes. 
In addition, it also included questions about time spent playing the game, the number of levels completed for each mini-game, and whether in-game quizzes were attempted.

Further, it included Likert-scale, ranking, and open-ended questions, based on \cite{wangenheim2012,morales-trujillo2018} questionnaires, to gather information on students' motivation, engagement, and overall experience with the mini-games.

\subsection{Data Collection and Analysis}

Learning outcomes were measured by comparing pre-game and post-game questionnaire scores. 
Additional behavioural measures, including time spent playing, number of levels completed, and completion of in-game quizzes, were examined in relation to post-game scores.

Engagement and motivation were assessed using self-reported Likert-scale responses, rankings of mini-games (based on engagement, motivation, and perceived teaching effectiveness), and indirect behavioural measures such as time spent playing and the number of levels completed. 

Descriptive statistics were used to summarise questionnaire responses, quiz scores, time spent playing, and levels completed. 
Pre-game and post-game questionnaire scores were compared to evaluate changes in learning outcomes. 
Relationships between engagement, motivation, and performance were examined descriptively to explore trends relevant to RQ1 and RQ2.

Open-ended responses were analysed using a lightweight thematic analysis approach. 
Data were coded inductively and grouped into recurring themes (e.g., visualisation, difficulty, guidance).

RQ1 was addressed using pre- and post-game scores to measure learning changes. 
RQ2 was examined using Likert-scale responses, rankings, and behavioural data (e.g., levels completed, quiz participation). 
These measures enabled analysis of relationships between engagement, motivation, and performance.

\section{Results}
\label{sec:results}

The preliminary survey in Phase One received 24 responses. 
Of these, 17 participants provided their email to be contacted for Phase Two, and 11 of those participants completed Phase Two of the study.

\subsection{Phase One Results}
\label{sec:phaseone-results}

Phase One involved a preliminary survey exploring learning preferences, game elements, and features that could support learning QC concepts in a game setting.

When asked whether quizzes or challenges within the game would be helpful for reinforcing concepts, 54\% of participants responded ``very helpful'', 38\% ``somewhat helpful'', 4\% ``neutral'', and 4\% ``not very helpful''.

Participants also ranked the importance of various game elements. 
Aesthetics and visual appeal received an average rating of 4.17 out of 5. 
Levels with increasing difficulty, a points or scoring system, and rewards were the three other highest-rated elements, with average scores of 4.29, 4.13, and 4.08, respectively. 
Narrative or storyline ranked lowest with an average score of 2.71, while badges or achievements and leaderboards also received relatively low scores (2.88 and 3, respectively).

When asked to rank features that would help with understanding complex concepts in games, participants ranked clear visualisations highest, followed by immediate feedback on actions, step-by-step instructions, and challenges that test understanding. 
Additionally, 83\% of participants preferred a balance between teaching and engaging gameplay.

Regarding preferred level duration, ``less than 5 minutes'' and ``5-10 minutes'' were tied at 42\% each. 
For motivational factors, achieving high scores and gaining a better understanding of the topic were the most commonly selected (67\%), followed by competing with others (54\%) and unlocking new content (38\%).

Responses to open-ended questions indicated common challenges, including games becoming \textit{``too abstract''} relative to the topic, levels being \textit{``too difficult''}, and a lack of guidance throughout the game. 
Suggestions for additional features included collectible items, visualisations, and cosmetic rewards such as unlockable outfits or backgrounds.

\subsection{Phase Two Results}
In Phase Two, 11 participants completed a pre-game questionnaire, played the mini-games, and completed a post-game questionnaire. 
The average pre-game questionnaire score was 2.73 out of 10, while the average post-game questionnaire score was 7.82 out of 10.

Four participants attempted all in-game quizzes and completed most or all levels of each mini-game, achieving an average post-game score of 9.5 out of 10, compared to an average pre-game score of 3.25 out of 10. 
Two participants completed all mini-game levels and all in-game quizzes, both achieving a post-game score of 10 out of 10, with an average pre-game score of 1.5 out of 10.

Participants reported that the game contributed positively to their learning, with one participant selecting ``a great deal'', four selecting ``a lot'', three selecting ``a moderate amount'', and three selecting ``a little''. 
The average time spent playing the game was 32 minutes, with times ranging from 10 to 60 minutes.

All participants played the Bloch sphere and entanglement mini-games. 
Seven participants played the quantum circuits mini-game, which was locked until completing six levels of the Bloch sphere mini-game. 
Each mini-game had a total of 12 levels, and on average, participants completed 9.6 levels of the Bloch sphere mini-game and 7.8 levels of the entanglement mini-game. 
Across all participants, the average number of quantum circuit levels completed was 5.5; among the seven participants who played this mini-game, the average was 8.4 levels.

\begin{figure}[!ht]
    \centering
\includegraphics[width=0.9\linewidth]{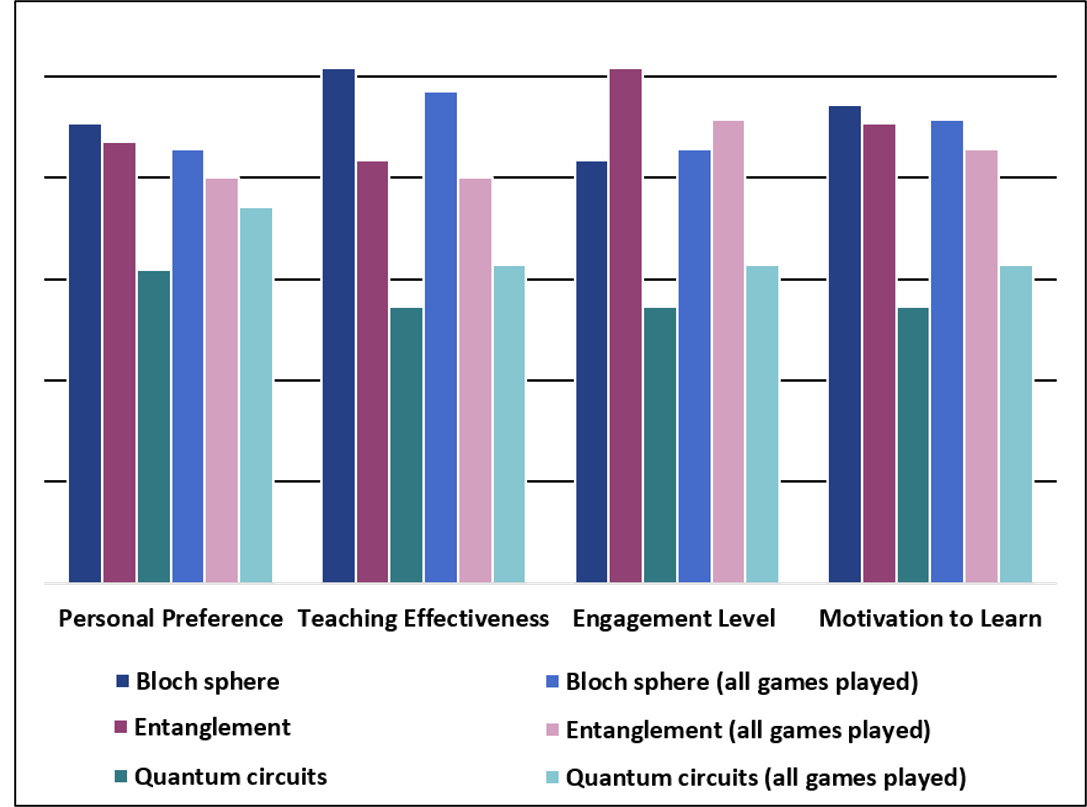}
    \caption{Ranking of the Mini-Games}
    \label{fig:games-ranking}
\end{figure}

Participants ranked the mini-games across four criteria: personal preference, perceived teaching effectiveness, engagement, and motivation. 
The Bloch sphere mini-game ranked first in personal preference, perceived teaching effectiveness, and motivation, while the entanglement mini-game ranked highest for engagement. 
The quantum circuits mini-game ranked lowest across all four criteria. 
Removing participants who did not play the quantum circuits mini-game did not change the ranking order. 
Figure~\ref{fig:games-ranking} shows the normalised rankings.

\begin{figure}[!ht]
    \centering
    \includegraphics[width=0.9\linewidth]{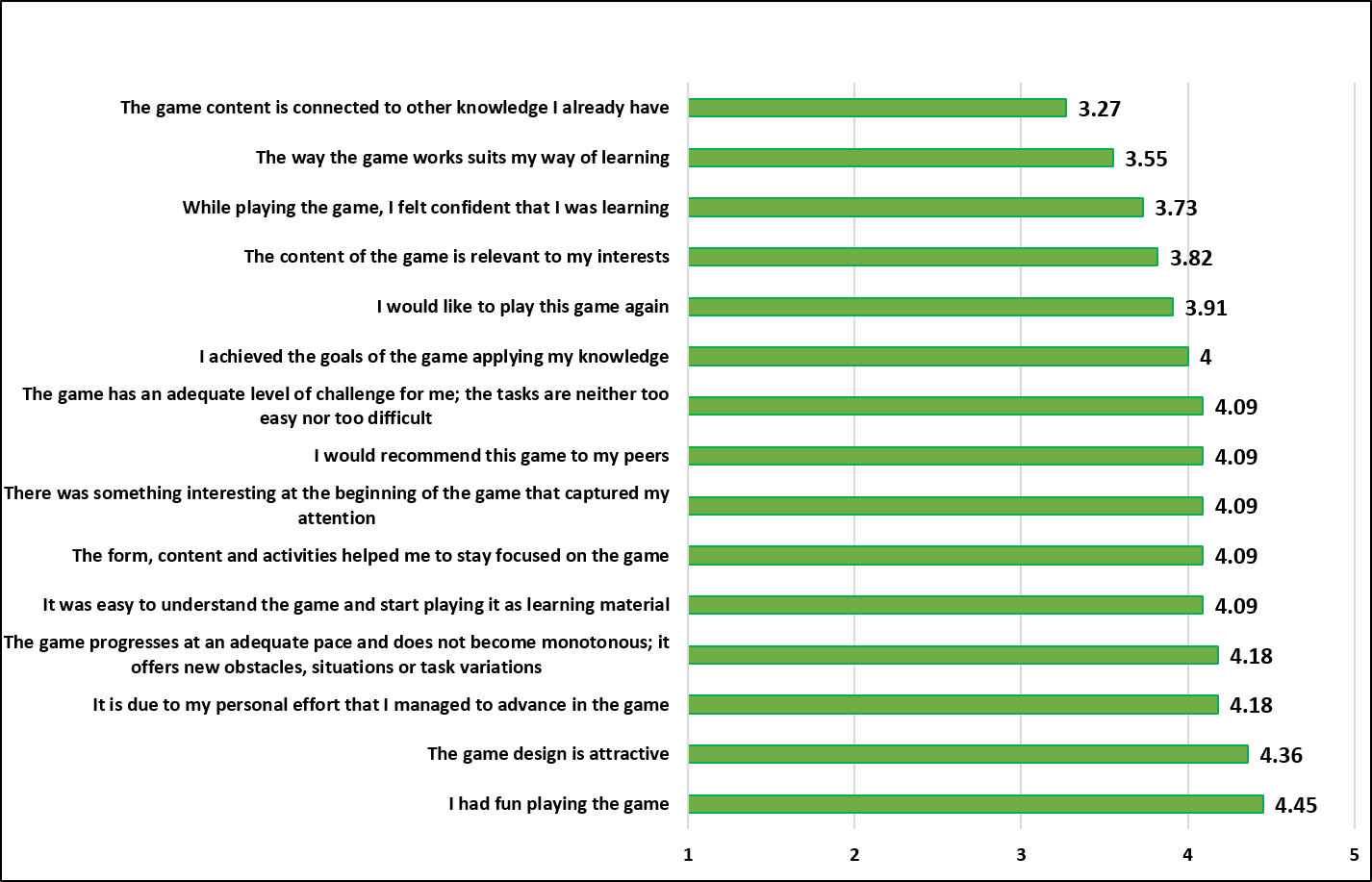}
    \caption{Motivation and Engagement Questions}
    \label{fig:engagement-questions}
\end{figure}

Responses to engagement and motivation questions indicated high overall engagement (see Figure \ref{fig:engagement-questions}). 
The highest-rated statements were ``I had fun playing the game'' and ``the game design is attractive'', with average scores of 4.45 and 4.36 out of 5, respectively. 
The lowest-rated statement was ``the game content is connected to other knowledge I already have'', with an average score of 3.27 (see Figure~\ref{fig:engagement-questions}).

Participants most commonly reported feeling challenged and interested while playing the game, with 10 and 9 participants selecting these feelings, respectively. 
No participants reported feeling anxious, sad, or angry, and only one participant reported feeling bored.

Nine participants completed at least one in-game quiz, with four completing all three. 
Of participants who completed at least one in-game quiz, 89\% reported that the quizzes were useful and that the difficulty level was appropriate.

Open-ended responses indicated that participants particularly enjoyed the visualisations, interactions with the Bloch sphere, and the game's overall theme. 
Common challenges included difficulty with the quantum circuits mini-game and feeling overwhelmed by some explanatory pop-ups. 
Suggested improvements included introducing game mechanics before QC concepts and providing more gradual explanations.

\section{Discussion}
\label{sec:discussion}

\subsection{RQ1: Effectiveness of Gamifying QC Concepts}

Although the sample size of participants in Phase Two was small, the results indicate that the game had a positive effect on learning outcomes. 
The average increase in correct answers between the pre-game and post-game questionnaires was 5.09, with participants improving their average score from 2.73 to 7.82 out of 10. 
This improvement was observed across all participants, as all participants achieved a higher score in the post-game questionnaire than in the pre-game questionnaire.

Participants who engaged more extensively with the game showed greater improvements. 
The four participants who completed nearly all levels of each mini-game and completed all in-game quizzes achieved an average increase of 6.25, resulting in an average post-game score of 9.5 out of 10. 
The two participants who completed all levels of each mini-game and all in-game quizzes both achieved perfect post-game scores of 10 out of 10, despite having very low pre-game scores, with an average increase of 8.5. 
While these results are based on a very small subset of participants, they suggest that deeper engagement with the game is associated with stronger learning outcomes.

It is also noted that some participants who spent only 10-15 minutes playing the game and completed relatively few levels still demonstrated improved post-game scores. 
This suggests that even limited exposure to the mini-games was sufficient to support learning improvements. 
Additionally, all participants reported that the game taught them something about QC, with 73\% indicating that it taught them a moderate amount or more. 
Taken together, these findings suggest that \textit{QubitQuest} successfully gamified introductory QC concepts in a way that supported learning and improved participants' understanding of the subject.

\subsection{RQ2: Motivation, Engagement, and Performance}

The results indicate a relationship between motivation, engagement, and performance, although this relationship is not conclusive due to the small sample size. 
As shown in Figure~\ref{fig:games-ranking}, the Bloch sphere mini-game ranked highest in terms of personal preference, perceived teaching effectiveness, and motivation to learn, while the entanglement mini-game ranked highest for engagement. 
The quantum circuits mini-game consistently ranked lowest across all criteria. These rankings were largely consistent even when excluding participants who did not play the quantum circuits mini-game, although the differences between rankings became smaller.

Because participants were asked to rank the mini-games relative to one another rather than rate each mini-game individually, the absolute level of motivation or engagement experienced by participants is difficult to quantify, being an improvement required for future studies.
Motivation and engagement varied substantially between participants and between mini-games. 
A more detailed assessment, such as asking participants to rate each mini-game on a numerical scale, would allow for clearer comparisons and conclusions.

Despite these limitations, a pattern was observed when comparing the mini-game each participant played the most with the mini-game they rated as the most motivating and engaging. 
Eight participants identified the mini-game they played the most as both the most motivating and the most engaging.
Two additional participants identified the mini-game they played the most as the most motivating, but not the most engaging. 
These results suggest a strong relationship between motivation and engagement, which aligns with previous works \cite{Schunk12, Aji18, Nand19}.

Interestingly, motivation appeared to be a slightly stronger factor than engagement in determining how much participants played a mini-game. 
This suggests that a mini-game's ability to motivate learning may be particularly important for encouraging sustained interaction. 
Given the relationship between the number of levels completed and post-game questionnaire scores, these findings indicate that increasing motivation and engagement can, indirectly, improve learning outcomes by encouraging participants to spend more time interacting with the game.

\subsection{Lessons Learned}
\label{sec:lessons-learned}

Based on the study results, several lessons were learned for the design and implementation of gamified educational tools for introductory QC concepts:

\begin{itemize}
    \item \textbf{Progressive levels and small learning units support learning.} Participants generally completed more levels in the mini-games that introduced one concept at a time and gradually increased in difficulty (i.e., Bloch Sphere and Entanglement). This may indicate that small steps and progressive levels can make learning less intimidating and students keep playing.

    \item \textbf{Interactive visualisations and feedback help learning and motivation.} Participants mentioned in open-ended responses and rankings that the Bloch sphere mini-game's interactive visualisations and immediate feedback made it easier to understand new and complex concepts (i.e., the quantum gates). These features also motivated them to keep playing.

    \item \textbf{Motivation affects how much time students spend playing.} Participants completed more levels in mini-games that they found motivating. This suggests that making mini-games motivating (e.g., with clear goals or progress) can encourage students to spend more time learning.

    \item \textbf{Optional in-game quizzes help learning and motivation.} Not all participants completed every in-game quiz, but those who did said they were useful and had the right difficulty. Quizzes give immediate feedback and can encourage students to try again, helping their learning and motivation.

    \item \textbf{Good design and visuals support engagement.} Participants rated the game design and visuals highly. Clear and attractive visuals helped students focus on learning and made them more willing to play.


    \item \textbf{Short and manageable levels may support engagement.} Most participants preferred levels shorter than 10 minutes in the preliminary survey, and the designed levels followed this constraint. 
    While this may support engagement, no controlled comparison of level length was conducted.
    
    \item \textbf{Mini-games should match students' prior knowledge.} The quantum circuits mini-game was played less, likely because it required more background knowledge. Matching difficulty to prior knowledge can prevent frustration and help students stay motivated.

    \item \textbf{Students can learn even without completing all levels.} Some mini-games had many levels, but participants who only completed part of them still improved their post-game questionnaire scores. This suggests that fewer levels may be ``enough'' to achieve the desired learning outcome, and reducing unnecessary levels could make the game more efficient.   
\end{itemize}

Gamified QC learning resources should prioritise progression, interactive visualisations, optional formative quizzes, and concise level design. 
Incorporating these elements can improve students' motivation, engagement, and performance.

\subsection{Limitations and Threats to Validity}
\label{sec:limitations}

Several limitations and threats to validity affect the interpretation of the results. 
The primary limitation of the study is the small sample size, with only 11 participants completing Phase Two. 
This limits the generalisability of the findings, as the results may not be representative of a broader population of CS/SE students. 
Offering incentives or integrating the study into a course context could help increase participation in future work.

The study did not include a control group, as the primary aim was to explore whether interaction with the game could improve learning outcomes and how engagement and motivation related to those outcomes. 
As a result, it is not possible to attribute learning gains exclusively to the game without considering other potential factors. 
Due to the small sample size, no inferential statistics were used, and the analysis relied on descriptive comparisons.

Another limitation was that not all participants completed all mini-games or levels. 
While this variation was useful for examining relationships between engagement, motivation, and learning outcomes, it reduced the number of participants who completed all content to only two. 
Future studies could benefit from a larger sample size to assess variation in completion rates across mini-games and, therefore, allow for meaningful comparisons.

Another threat to validity relates to the measurement of motivation and engagement. 
The questionnaire items were intentionally kept simple to minimise participant burden, but this limited the depth of insight into these constructs. 
A more comprehensive questionnaire, combined with qualitative methods such as interviews, could provide a richer understanding of how motivation and engagement influence learning.

Finally, learning outcomes were assessed using multiple-choice quizzes. 
While this allowed for straightforward quantitative analysis, such assessments may not fully capture a deeper understanding of QC concepts. 
A more robust evaluation could include formal assessment items within a course setting to validate learning alongside the current measures.
This also could mitigate the potential threat present here, where participants volunteered for the study, which may bias the sample towards students already interested in QC or games in general.

\section{Conclusions and Future Work}
\label{sec:conclusion}

This paper presented \textit{QubitQuest}, a set of mini-games designed to teach introductory QC concepts to CS/SE undergraduates. 
The goal was to gamify QC concepts to support learning while also being engaging and motivating. 
Existing QC games and related work were analysed to guide the design of the proposed solution.

Three QC mini-games were developed, and a user study was conducted in which participants played the mini-games and completed questionnaires and quizzes to assess their performance, motivation, and engagement. 
The results showed that participants improved their understanding of introductory QC concepts after playing the mini-games. 
In addition, participants who completed more levels achieved higher post-game questionnaire scores. 
Motivation and engagement also influenced how much of the mini-game participants played, which, consequently, affected their learning outcomes.

The results highlighted several directions for future work. 
In the current version of the game, design decisions such as level locking and level complexity were based on the authors' experience. 
Future versions could adjust these elements using participant feedback; for example, by simplifying explanations, reducing the number of levels where appropriate, or changing when levels are unlocked. 
Additional improvements could also be made after the game is tested by a larger group of students.

Another future goal is to develop two additional mini-games, covering measurement and probability in more depth. 
The current game architecture already supports adding new mini-games, making this a practical extension. 
One possible direction is to include a competitive mini-game, as some participants indicated interest in competitive elements during the study.

Finally, a future goal is to run a larger user study with more participants and to include interviews. 
The mini-games could also be integrated into a QC course alongside traditional teaching, which would allow learning outcomes to be studied in a more realistic educational setting.

\section*{Acknowledgments}
The authors thank students for participating in the study and technical staff for providing the infrastructure to run the application.

\balance

\bibliographystyle{ACM-Reference-Format}
\bibliography{references}

\end{document}